\documentclass[final]{svjour3}
\usepackage{graphicx}
\usepackage{rotating}
\usepackage{amssymb}
\usepackage{siunitx}
\usepackage{physics}
\usepackage{}
\usepackage{bm}
\usepackage{mathptmx}
\usepackage{wrapfig,booktabs,tabularx,caption}
\usepackage[hyphens]{url}
\makeatletter
\journalname{Journal of Low Temperature Physics}
\begin{document}
\raggedbottom
\setlength{\belowcaptionskip}{-10pt}
\setlength{\floatsep}{5pt}
\setlength{\intextsep}{5pt}
\setlength{\textfloatsep}{5pt}
\newcommand{\hdblarrow}{H\makebox[0.9ex][l]{$\downdownarrows$}-}

\title{First Calorimetric Measurement of Electron Capture in ${}^{193}$Pt with a Transition Edge Sensor}

\author{K.E.~Koehler \and M.A.~Famiano \and C.J.~Fontes \and T.W.~Gorczyca \and M.W.~Rabin \and D.R.~Schmidt \and J.N.~Ullom \and M.P.~Croce}

\institute{K.E. Koehler \and M.P. Croce \and C.J. Fontes \and M.W. Rabin\\Los Alamos National Laboratory\\Los Alamos, NM 87545, USA\\Tel.: +1 (505) 695 4100\\
\email{kkoehler@lanl.gov}\\\\K.E. Koehler \and M.A. Famiano \and T.W. Gorczyca\\Western Michigan University\\Kalamazoo, MI 49008, USA\\\\D.R. Schmidt \and J.N. Ullom\\National Institute of Standards and Technology (NIST)\\Boulder, CO 80305, USA}

\maketitle

\begin{abstract}

The neutrino mass can be extracted from a high statistics, high resolution calorimetric spectrum of electron capture in ${}^{163}$Ho. In order to better understand the shape of the calorimetric electron capture spectrum, a second isotope was measured with a close to ideal absorber-source configuration. ${}^{193}$Pt was created by irradiating a ${}^{192}$Pt-enriched platinum foil in a nuclear reactor. This Pt-in-Pt absorber was designed to have a nearly ideal absorber-source configuration. The measured ${}^{193}$Pt calorimetric electron-capture spectrum provides an independent check on the corresponding theoretical calculations, which have thus far been compared only for ${}^{163}$Ho. The first experimental and theoretically-calculated spectra from this ${}^{193}$Pt-in-Pt absorber are presented and overlaid for preliminary comparison of theory with experiment.

\keywords{neutrino mass, electron capture, microcalorimeter, superconducting transition-edge sensor}

\end{abstract}

\section{Introduction}

Research to understand the calorimetric measurement of electron capture and beta decay within a low temperature detector is of interest for neutrino science and metrology~\cite{Nucciotti2016,Hassel2016,Faverzani2016,Croce2016,Rotzinger2008,Loidl2008,Loidl2010,Loidl2017}. A neutrino mass measurement from microcalorimeter spectroscopy of electron capture decay in ${}^{163}$Ho requires a large-scale experiment with stringent requirements on statistics and detector resolution and a validated theoretical model of the electron capture spectrum. For example, an experiment with $\Delta E=1$~eV FWHM, pulse pile up fraction ($f_{pp}$) of $10^{-6}$, and a spectrum of $10^{16}$ decays can reach a $0.2$~eV neutrino mass sensitivity with $90$\% confidence level \cite{Nucciotti2016}. As a tool for understanding the theoretical shape of calorimetric measurements of electron capture, experimental measurements of ${}^{193}$Pt were made and compared to theoretical calculations. These are the first published results of either experimental measurements or theoretical calculations of the ${}^{193}$Pt calorimetric electron capture spectroscopy.

${}^{163}$Ho with its low $Q$ value of 2.833 keV is far more desirable for a neutrino mass measurement because the statistics in the endpoint region increase with lower $Q$; however, ${}^{193}$Pt has the next lowest $Q$ value for electron capture of 56.8~keV and a reasonable half life of 50 years~\cite{Basunia2017}. Significant work has been done in the field on improving the theoretical calculations for the electron-capture spectrum \cite{DeRujula2013,Faessler2015-1hole,Robertson2015_objection,Faessler2015-2hole,Faessler2015-3hole,DeRujula2015,DeRujula2016,Faessler2016a,Faessler2017}, but all comparisons to date have been based on ${}^{163}$Ho, with some early work on internal bremsstrahlung electron capture with ${}^{193}$Pt~\cite{DeRujula1981,Jonson1983}. Comparing to another isotope is a valuable check on the universal applicability of these models and ensures that the community does not begin to rely on isotope-specific approximations or empirical scaling, which might bias the neutrino mass results.

%Previous microcalorimeter measurements by the Los Alamos National Laboratory low temperature detector team with superconducting transition edge sensors (TESs) have shown a degraded energy resolution in measured ${}^{163}$Ho spectra~\cite{Croce2016}.  The energy resolution of a calorimetric spectrum of ${}^{163}$Ho, a solution dried in a nanoporous gold absorber, was shown to be significantly worse than the spectrum from an electroplated ${}^{55}$Fe source, another isotope that decays via electron capture~\cite{Croce2016}. This degradation is quantified by fitting the data to a Bortels function, the convolution of a Gaussian with an exponential low energy tail~\cite{Bortels1987}. The width of the Gaussian component ($\Delta E = 2.355 \sigma$) is $\Delta E = 43$~eV (M1 peak) for the ${}^{163}$Ho spectrum, but only $\Delta E = 7.5$~eV for the ${}^{55}$Fe spectrum. Similarly, the tailing factor ($\tau$) from the exponential component is also higher in the ${}^{163}$Ho spectrum ($\tau=28$~eV for the M1 peak) compared to the ${}^{55}$Fe spectrum ($\tau=10$~eV), indicating more incomplete energy-thermalization in the absorber with ${}^{163}$Ho. It is hypothesized that this effect is due to the material co-deposited on the absorber, when the ${}^{163}$Ho-containing solution is dried onto the absorber \cite{Croce2016}. Additionally, we can rule out a specific effect of ${}^{163}$Ho because the Electron Capture ${}^{163}$Holmium collaboration (ECHo) has achieved a $\Delta E = 8.3$~eV resolution~\cite{Ranitzsch2014}.

A ${}^{193}$Pt-in-Pt absorber was created by neutron-irradiating a ${}^{192}$Pt-enriched foil, to create a source-absorber matrix as close to ideal as possible. This absorber should have no complex matrix structure in the form of additional material from dried deposits~\cite{Croce2016} or lattice damage from ion implantation (the route pursued by ECHo \cite{Gastaldo2014} and HOLMES \cite{Alpert2015} for embedding ${}^{163}$Ho in a gold foil). A small piece of the resulting irradiated foil can then be used as the absorber material with minimal additional sample preparation. This source-absorber configuration is ideal because the ${}^{193}$Pt is produced homogeneously within the Pt host and there are no elemental interfaces between the decaying isotope and the surrounding absorber material, which is a high purity metal. The absorber should have a simple metallic behaviour at low temperature. The heat capacity and thermal conductivity are well-understood and the absorber structure is expected to be nearly perfect, with fewer opportunities for energy trapping and a uniform environment for energy deposition and thermalization.

\section{Irradiation and Characterizations}

A 10~mg sample of 56.9\% ${}^{192}$Pt-enriched Pt was irradiated at the 6-MW thermal research reactor at the Massachusetts Institute of Technology for approximately 7 days. The original platinum foil had 675~ppm of total contaminants~\cite{PtAssay}, with the majority being lead (330~ppm) and iron (210~ppm). All other contaminants were $\leq20$~ppm. These impurities may contribute extra heat capacity. The platinum foil was placed in a polyethylene tube and sent through the 1PH1 intermediate flux pneumatic tube and irradiated at a thermal power of 5.74 MW for 7 days.

The activity of the created ${}^{193}$Pt was estimated in two ways. In the first method, using a 50-year half life~\cite{Basunia2017}, a 10 barn cross section\footnote{An experimental measurement of the cross section at the single energy of 0.025~eV yielded $10\pm2.5$ barns~\cite{Vertebnyi1975}.} for thermal neutron capture on ${}^{192}$Pt, an estimated thermal neutron flux of $8 \times 10^{12}$ n/cm${}^2$-s, and an irradiation of 7~days, we will produce 43~Bq of ${}^{193}$Pt per \si{\micro\gram} of platinum. The second method measured the ${}^{193\text{m}}$Pt content by gamma spectroscopy using an HPGe detector, giving a lower bound on ${}^{193}$Pt of 0.75~Bq/\si{\micro\gram}. The most active isotope determined through gamma spectroscopy was ${}^{192}$Ir at 18.6~Bq/\si{\micro\gram}, despite Ir constituting $<0.8$~ppm from an assay of the original foil. The most dominant contaminants apart from ${}^{192}$Ir were ${}^{51}$Cr and ${}^{182}$Ta with specific activities of 0.06~Bq/\si{\micro\gram} and 0.08~Bq/\si{\micro\gram} respectively. These contaminants each have less than a 1-year half-life and will decay away.

\section{Theoretical Spectrum}
\label{sec:theory}

The simplest model of electron capture involves a proton within the nucleus capturing an orbital electron, leaving a single vacancy in the daughter atom. As a result of the vacancy, the daughter is now in an excited atomic state. Embedding this daughter within an absorber allows a calorimetric measurement of all the decay energy except that of the escaping neutrino and potential surface escapes of photons or electrons. The equation describing the calorimetrically-measured electron capture spectrum comes from Fermi's description of $\beta$-decay~\cite{Fermi1934,DeRujula1981,DeRujula1982}:
\begin{align}
\label{eq:dgammade}
\frac{d \Gamma_{TOT}}{d E_c} \propto (Q - E_c) \sqrt{\left(Q-E_c\right)^2-{m_\nu}^2} \sum_{f} G^2 \zeta \frac{W_f \frac{\Gamma_{f}}{2\pi}}{(E_c-E_{f})^2+\frac{\Gamma_{f}^2}{4}}.
\end{align}
The total decay rate is $\Gamma_{TOT}$ and $E_c$ is the measured calorimetric energy. The neutrino density of states contributes the prefactor before the sum, dependent on the energy released by the reaction ($Q$) and the neutrino kinematic mass ($m_\nu$). $G$ is the weak Fermi coupling factor and $\zeta$ is the nuclear matrix element. The sum over final states is a sum over all allowed post-nuclear-decay states in the daughter. The weighting of each Lorentzian ($W_f$) depends on both the electron-nucleus interaction and the parent-daughter atomic wavefunction overlap. The hard work is in determining which post-decay states to keep and calculating the weighting factor.

Using antisymmetrized wavefunctions and Faessler's notation~\cite{Faessler2015-1hole,Faessler2015-2hole,Faessler2015-3hole,Faessler2016a,Faessler2017}, $W_f$ can be approximated as:
\begin{align}
\label{eq:faessler}
W_f \approx \left|\sum_{i=1\ldots Z} \left( \delta_{fi} \prod_{k \neq f} \braket{k'}{k} - \delta_{f \neq i} \braket{i'}{f} \prod_{k \neq f,i } \braket{k'}{k} \right)\Psi_i(R)\right|^2,
\end{align}
where $\ket{k}$ is an orbital calculated in the parent nuclear and atomic potential and $\ket{k'}$ is an orbital calculated in the daughter nuclear and atomic potential. $Z$ is the atomic number of the parent atom. $\Psi(R)$ represents the electron-nucleus interaction, here described by an orbital wavefunction in the parent evaluated at the nuclear radius, $R$. The simplest approximation which we refer to as the unitary approximation is to assume $\braket{i'}{j}=\delta_{ij}$, so $W_f$ simplifies to $|\Psi_f(R)|^2$.

There are various corrections to this simple model to be explored by comparisons to experimental data. A few of the corrections follow: (1) the daughter state may involve not only a vacancy from an electron capture, but an additional electron (or more) may be promoted to another bound state (shake up) or to the continuum (shake off)~\cite{Robertson2015_objection,Faessler2015-2hole,Faessler2015-3hole,DeRujula2015,DeRujula2016,Faessler2016a,Faessler2017}; (2) the atomic overlap can be calculated using a product of single-electron wavefunction overlaps as shown in Eq.~\ref{eq:faessler}, rather than assuming the unitary approximation; (3) more terms from the fully anti-symmetrized atomic wavefunction can be included, increasing the complexity and accuracy of an atomic overlap calculation. This paper does not address (1), but does show the spectroscopic effects of approximating the atomic overlap as unity versus a calculated atomic overlap (2) and the effects of including more terms in the atomic overlap (3).

\begin{wraptable}{l}{3.2cm}
\caption{Values for the binding energies and line widths given for the calculated ${}^{193}$Pt spectra from the literature~\cite{Campbell2001,CRC}. Values marked with an asterisk have not been measured and were used as placeholders.\label{tab:Pt}}
\vspace{0.15in}
%\centering
\begin{tabular}{@{}lrr@{}}
\toprule
Peak & $E$ (eV) & $\Gamma$ (eV) \\
\midrule
L1	 & 13419 & 7.9	 \\
L2	 & 12824 & 5.23	 \\
M1	 & 3174	 & 14.8	 \\
M2	 & 2909	 & 8.9	 \\
N1	 & 691.1 & 8	   \\
N2	 & 577.8 & 6.1	 \\
O1	 & 95.2	 & 5*	   \\
O2	 & 63	   & 5*	   \\
P1	 & 1*	   & 1*	   \\
\bottomrule
\end{tabular}
\end{wraptable}

Allowed captures are governed by conservation of energy and angular momentum, so the higher $Q$ value for electron capture in ${}^{193}$Pt allows captures from the ns and np${}_{1/2}$ states, where n $\geq$ 2, whereas the lowest states accessible for electron capture in ${}^{163}$Ho are the 3s and 3p${}_{1/2}$. The energies and widths of the ${}^{193}$Pt peaks from \cite{Campbell2001,CRC} are given in Table~\ref{tab:Pt}.

The nomenclature describing the calculated spectra is 1H-$\mathcal{O}$(N), where 1H denotes a single hole spectrum (no shake-up or shake-off contributions). N describes the number of off-diagonal orbital overlap factors retained in the calculation. In principle, this description indicates which terms are being kept from the overlap of the fully-antisymmetrized atomic wavefunctions, each of which has Z! terms. For example, if only the first term in Eq.~\ref{eq:faessler} is kept, this description is $\mathcal{O}$(0) because all orbital overlaps are of type $\braket{k'}{k}$. If all the written terms in Eq.~\ref{eq:faessler} are kept, this description is $\mathcal{O}$(1) because we are keeping terms that have one off-diagonal factor, $\braket{i'}{f}$. The unitary approximation described above ($\braket{i'}{j}=\delta_{ij}$) is referred to as $\mathcal{O}$(0U).

Both 1H-$\mathcal{O}$(0U)a and 1H-$\mathcal{O}$(0U)b are similar calculations with the only difference coming from the evaluation of the orbital wavefunction at the nucleus. Model ``a'' uses tabulated values for the orbital Dirac-Fock-Slater (DFS) wavefunctions at $r = 0$~\cite{Band1986}, whereas Model ``b'' uses uses orbital wavefunctions from the LANL atomic code suite~\cite{Fontes2015} evaluated at $r = 1.365 \times 10^{-4}$ au, a good approximation for the ${}^{193}$Pt nuclear radius~\cite{Angeli2013}. The other two models calculated also use orbital wavefunctions from the LANL atomic code suite for use in the atomic overlap calculations. The biggest difference comes from the difference between Model ``a'' and Model ``b'', indicating the sensitivity to the orbital overlap with the nucleus. There are small differences in the predicted peak heights from the 1H-$\mathcal{O}$(0U)b, 1H-$\mathcal{O}$(0), and 1H-$\mathcal{O}$(1) theories. These theoretical curves are all shown in Fig.~\ref{fig:spectrum} alongside experimental data.

\section{Experimental Calorimetric Spectrum}

\begin{figure}[htbp]
\begin{center}
\includegraphics[width=\linewidth]{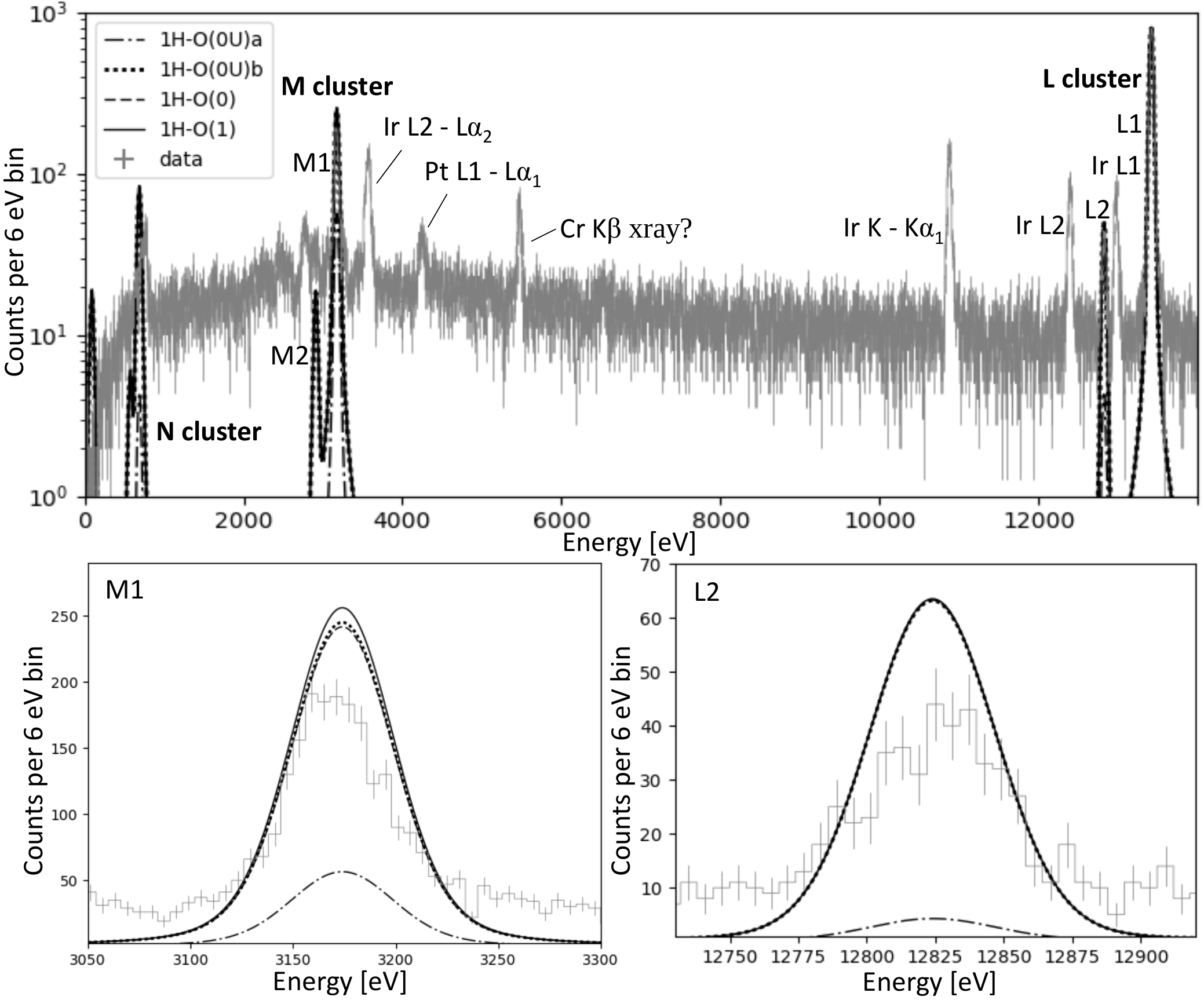}
\caption{(\textit{Top}) Calorimetric measurement of ${}^{193}$Pt (grey) including contaminants such as ${}^{192}$Ir compared to one-hole excitation theoretical curves for ${}^{193}$Pt (black). Experimental data is single-point energy-calibrated with the L1 peak. Theory curves are convolved with a $\sigma=22$~eV Gaussian and are scaled to the L1 peak height. A single-point energy calibration on L1 has only a slight mismatch for the M1 peak (\textit{Bottom Left}). \textit{Bottom Left} and \textit{Bottom Right} show inadequacy of the 1H-$\mathcal{O}$(0U)a model to predict the M1 and L2 peak heights, respectively. More statistics are needed to discriminate between the other three theories, all of which appear to overshoot the M1 and L2 peak heights. See text for description of nomenclature.\label{fig:spectrum}}
\end{center}
\end{figure}

The transition-edge-sensor microcalorimeter is a 350~\si{\micro\meter} square Mo-Cu bilayer with a superconducting transition temperature near 110 mK~\cite{Croce2016}. A small piece ($\approx 20 \times 20 \times 5$~\si{\micro\meter}${}^3$ or 0.04~\si{\micro\gram}, estimated from density) from the irradiated foil was hand-cut with a razor and attached to the microcalorimeter detector with an indium bump bond. The preliminary measured spectrum (Fig.~\ref{fig:spectrum}) employed a linear energy calibration through the origin using the L1 capture peak. Future measurements will use external calibration lines to ascertain the shift in the peaks, since recent discussion has indicated that the energies used should not be just the binding energy of the daughter orbitals, but the energy of the daughter configuration~\cite{Nucciotti2016,Springer1985,Robertson2015}. For ${}^{163}$Ho, this amounts to adding the binding energy of the additional electron in the 4f shell. For ${}^{193}$Pt, this is not quite as straightforward because the ground state of Ir and Pt involves a rearrangement of both the 6s and 5d shells. While Pt has 9 electrons in the 5d shell and 1 electron in the 6s shell, the ground state of Ir has 7 electrons in the 5d shell and 2 electrons in the 6s shell. The energy of the Lorentzian peak center might then be calculated as $\text{BE}_{f}+2\text{BE}_{5\text{d}}-\text{BE}_{6\text{s}}$, where the $f$ denotes the vacancy orbital and the binding energies are with respect to the Ir atom. Because of this difference, until a higher statistics and higher resolution measurement can be made, we are using a very simple energy calibration.

While the production of ${}^{192}$Ir was unintentional, the spectrum does show the peak signature of electron capture in ${}^{193}$Pt in addition to ${}^{192}$Ir, which has a 4\% decay path via electron capture. The data shown in Fig.~\ref{fig:spectrum} represents a 58-hour measurement along with scaled theoretical descriptions convolved with a $\sigma = 22$~eV Gaussian response. No attempt has been made to fit the theoretical descriptions to the data. A simple scaling of the calculated theoretical shapes to the L1 peak was done and an adequate Gaussian description was determined by varying $\sigma$ until it was a reasonable description of the data. 

The dominant peak (L1) comes from 2s capture in ${}^{193}$Pt, with the next three peaks descending in energy coming from 2s capture in ${}^{192}$Ir (L1), 2p${}_{1/2}$ capture in ${}^{193}$Pt (L2), and 2p${}_{1/2}$ capture in ${}^{192}$Ir (L2). The keen observer will notice that the ${}^{192}$Ir L2 peak is actually taller than the ${}^{192}$Ir L1 peak, contrary to expectations. This is because the ${}^{192}$Ir L2 peak is also populated by 1s captures in ${}^{192}$Ir (at 74 keV and out of dynamic range for this detector), but with an escape of the K$\alpha_2$ x-rays. The next peak at 10.9~keV is also from the 1s capture in ${}^{192}$Ir, but with the K$\alpha_1$ escape x-ray. The three peaks at around 3544~eV, 4241~eV, and 5427~eV are tentatively identified as an escape peak from ${}^{192}$Ir L2 peak, escape peak from ${}^{193}$Pt L1 peak, and a Cr x-ray. The M cluster contains M1 and M2 peaks for both Pt and Ir, similarly the N cluster. Higher statistics are needed to better identify the peaks in the M and N clusters. Further measurements will also help in identification of the peaks between the M and L clusters because ${}^{192}$Ir has a much shorter half life (74 days) than ${}^{193}$Pt (50 years), so it will disappear from the spectrum over time. 

A lower bound for the activity of the ${}^{193}$Pt was calculated by integrating the experimental data in the L1, L2, and M1 peaks above background (a flat background assumed for each peak) for an activity of $0.0488 \pm 0.0006$ Bq (1.2 Bq/\si{\micro\gram}). A plausible actual activity is calculated by integrating the scaled theoretical fit for 1H-$\mathcal{O}$(1) and dividing those counts by the measurement time. Based on these calculations, the ${}^{193}$Pt activity within the absorber was 0.13 Bq (3.3 Bq/\si{\micro\gram}), larger than the lower bound provided by gamma spectroscopy measurements of ${}^{193\text{m}}$Pt (0.75 Bq/\si{\micro\gram}). This calculation is an order of magnitude lower than the predicted specific activity from a 10-barn cross section measurement (41 Bq/\si{\micro\gram}), but reasonable assuming a 1-barn cross section calculation (4.1  Bq/\si{\micro\gram}), which might indicate the neutron flux was not as thermal as expected. The lower bound given by integrating the L1, L2, and M2 experimental peaks is the most conservative conclusion and indicates the 10-barn cross-section calculation is deficient in the values used for integrated neutron flux, neutron energies, and/or cross-section.

Preliminary comparisons of experimental data to theoretical curves are encouraging. All binding energies and line widths used in the theoretical calculations come from the literature~\cite{Campbell2001,CRC} (see Table \ref{tab:Pt}), so the only empirically derived values were an overall scaling factor determined from the L1 peak and $\sigma$. The locations and approximate heights of the L2 and M1 peaks on the ${}^{193}$Pt spectrum are accurately described by the theory. The L2 and M1 peak heights are overshot by the theoretical curves, apart from the 1H-$\mathcal{O}$(0U)a theory. There are several potential factors at play here: the detector is not a perfect $4\pi$ detector because the irradiation of the foil would have created ${}^{193}$Pt on the edges of the foil; the background is substantial and there is a significant background in the spectrum; and the peak heights are strongly dependent on the nuclear radius used and the shape of the calculated atomic orbital near the origin because the Lorentzian peaks scale like $|\Psi(R)|^2$. Any further comparisons of theoretical shape and experimental data would be premature given the low statistics of the spectrum presented.

\section{Conclusions and Future Work}

This paper presents the first calorimetric measurement of electron capture in ${}^{193}$Pt and the first published theoretical calculations for this spectrum. It is clear that further research must be done on the detector response ($\Delta E= 52$~eV FWHM). The predicted resolving power of 290 based purely on signal and noise considerations corresponds to 46~eV FWHM at the L1 peak (13419~eV). This is consistent with the measured detector response, indicating the energy resolution degradation may be due to heat capacity. However, while this is a less than satisfactory energy resolution, the spectrum does not show a significant difference in resolution between the L and M clusters, such as that seen between the M and N clusters in previously measured ${}^{163}$Ho spectra by this team~\cite{Croce2016}. Better identification of the peaks between the L and M cluster should also be pursued. With higher statistics and/or improved resolution the various theoretical curves can be compared to the experimental data, offering an independent check on the theory from the work done with ${}^{163}$Ho.

\pagebreak

\begin{acknowledgements}
This work was supported by the US Department of Energy (DOE) Nuclear Energy's Fuel Cycle Research and Development (FCR\&D), Materials Protection, Accounting and Control Technologies (MPACT) Campaign and Los Alamos National Laboratory, Laboratory-Directed Research and Development Program. We gratefully acknowledge the support of the Center for Integrated Nanotechnologies, an Office of Science User Facility, and the Massachusetts Institute of Technology reactor personnel, in particular Thomas Bork for facilitating the radiation and Mike Ames for conducting immediate gamma measurements and modeling the irradiation. Heartfelt thanks to Dave Mercer for his assistance with the gamma spectroscopy and analysis, and Andrew Hoover for participating in peer review of this work.
\end{acknowledgements}

\bibliography{bibliography}

\end{document}